 \documentclass[10pt, conference, twocolumns]{IEEEtran}
\usepackage{setspace}
 \usepackage[fleqn]{amsmath}

  \usepackage{hyperref}
  \hypersetup{
    pdfborder = {0 0 0},
    colorlinks = true,
    linkcolor = black,
    anchorcolor = red,
    citecolor = blue,
    filecolor = red,
    pagecolor = red,
    urlcolor = red
}
\usepackage{color}
\usepackage{polpack}
\usepackage{epstopdf}
\usepackage{xfrac}
\usepackage{bm}
\newcommand{\myemph}[1]{#1\xspace}
\IEEEoverridecommandlockouts

\begin{document}
\title{Learning-Based Optimization of Cache Content in a Small Cell Base Station}

\author{\IEEEauthorblockN{Pol Blasco and Deniz G{\"u}nd{\"u}z}
\IEEEauthorblockA{Imperial College London,  UK\\
Emails: \{p.blasco-moreno12, d.gunduz\}@imperial.ac.uk}
\vspace{-1cm}
}

\maketitle

\begin{abstract}Optimal cache content placement in a wireless small cell base station (sBS) with limited backhaul capacity is studied. The sBS has a large cache memory and provides content-level selective offloading by delivering high data rate contents to users in its coverage area. The goal of the sBS content controller (CC) is to store the most popular contents in the sBS cache memory such that the maximum amount of data can be fetched directly form the sBS, not relying on the limited backhaul resources during peak traffic periods. If the popularity profile is known in advance, the problem reduces to a \textit{knapsack problem}. However, it is assumed in this work that, the popularity profile of the files is not known by the CC, and it can only observe the instantaneous demand for the cached content. Hence, the cache content placement is optimised based on the demand history. By refreshing the cache content at regular time intervals, the CC tries to learn the popularity profile, while exploiting the limited cache capacity in the best way possible. Three algorithms are studied for this cache content placement problem, leading to different \textit{exploitation-exploration trade-offs}. We provide extensive numerical simulations in order to study the time-evolution of these algorithms, and the impact of the system parameters, such as the number of files, the number of users, the cache size, and the skewness of the popularity profile, on the performance. It is shown that the proposed algorithms quickly learn the popularity profile for a wide range of system parameters.
\end{abstract}

\section{Introduction}

Downlink traffic in cellular networks has been growing rapidly. It is envisioned that, with the increasing demand for high data rate and delay intolerant services, such as video streaming, the exponential growth in mobile downlink traffic will continue in the foreseeable future. Network densification, by deploying an increasing number of small cells (i.e., micro-, pico-, and femtocells), is considered to be the only viable option that can boost the capacity of  cellular networks at a scale that can match the growing demand.   Small cells, which are deployed by either the service provider or by a third party (e.g., private institutions and users), boost wireless capacity in areas with low coverage and/or high traffic demand such as, buildings and metro stations, and, most importantly, offload traffic from the rest of the macro cellular network. In a cellular network, each base station (BS) has a reliable high-capacity backhaul link to the core network, which enjoys high throughput and low delay. However, in dense networks, due to physical and cost-related limitations, small cell BSs (sBS) are connected to the core network through low-capacity and unreliable backhaul links.

Caching content at sBSs in order to increase the quality of experience (QoE) of the users and alleviate congestion in the sBS backhaul connection has received great attention from both the industry, with the development of cache-enabled sBSs \cite{UbiquisysWebpage[online]2013,SagunaWebpage[online]2013}, and the academia \cite{caching:Dai2012,caching:Golrezaei2012a,caching:Poularakis2013}.  Enabled by the drastic reduction in price and size of data-storage devices, significant amount of popular content, such as news feeds or YouTube videos, can be stored in sBSs, and reliably delivered to users when requested, without consuming the scarce bandwidth of the backhaul connection.

Considering the huge number of available content with varying size and popularity, an important problem is to decide which content should be cached in the limited storage space available \myemph{at the sBS. Cache content placement has been studied  for different cache topologies and applications. In  \cite{caching:S.Borst2010}, a cache cluster formed by several leaf-cache nodes and a parent-cache node is studied. Files requested by users connected to a leaf-cache node are fetched either directly from the leaf-cache node, from another cache node, or from the core network. The problem of optimally placing content in  the cache nodes in order to minimise the total bandwidth consumption is studied, and approximate solutions  are given for special cases.} The cache cluster in \cite{caching:S.Borst2010} is further studied in \cite{caching:Poularakis2012} for the case of equal size files and when leaf-cache nodes are not inter-connected, and a polynomial complexity optimal algorithm is given for the case of two leaf-cache nodes.   A caching problem with several caches in a dense wireless network is addressed in \cite{caching:Dai2012}.  Each cache aims at maximising the QoE of the users that are connected to it by allocating some of the available bandwidth to serve its users, and trading the rest with other caches in an auction-game fashion. Reference  \cite{ caching:Golrezaei2012a} studies a caching problem in a backhaul-constrained small cell network, in which users can connect to several sBSs, and fetch files from the sBS that offers the least latency. The problem is shown to be NP-hard, and approximate algorithms are given. 
In \cite{caching:Poularakis2013} users move randomly across an sBS wireless network.  At each time slot, users can access a single sBS, and download only a part of a content from the sBS cache. Coded content caching in the sBSs is optimised such that the amount of users that fetch the entire content file from the sBSs is maximised.  

References  \cite{caching:S.Borst2010,caching:Poularakis2012,caching:Dai2012,caching:Golrezaei2012a,caching:Poularakis2013} assume that either files' instantaneous demands (i.e, the number of requests per file), or files' popularity profile (i.e., the expected number of requests per file) are known noncausally. In this paper we take a more practically relevant approach, and assume that neither the instantaneous demands nor the popularity profiles are known in advance. Instead, assuming stationary file popularity (file popularity can be assumed stationary over short time spans, for example within a day), we derive algorithms that learn the best caching strategy by observing the instantaneous demands over time.

We study the optimal content caching problem in a private wireless sBS with limited backhaul. The sBS provides high data rate service to its users, and the rest of the traffic is offloaded to the macro cellular network. This may stand for the case of an sBS owned by the local transport authority  and located in a crowded metro station, or an sBS deployed in the facilities of a  private institution to provide service to its members. The sBS is equipped with a large cache memory, where popular content can be stored, and the sBS content controller (CC) is in charge of managing the content of the cache. Periodically, the sBS broadcasts information about the content of the cache (i.e., a list of the stored files) to its users. When a user wants to access high data rate content, if this content is included in the most recent cache content list, a request is sent to the sBS and the content is readily delivered to the user. If the content is not located in the cache, due to the scarce backhaul capacity of the sBS, no request is send to the sBS, and it is downloaded from the macro BS. The use of the broadcast signal avoids locking the sBS with requests that can not be served, and it can be completely  transparent to the users; for example, managed by a smart phone application that runs in the background, listens to the sBS broadcast signals, and sends user's requests either to the sBS or the to macro BS depending on the cache content. This, we call \emph{content-level selective offloading}.  

The  objective of the CC is to find the best set of files to cache in order to maximise the traffic served from the sBS  without knowing the popularity profile in advance, and by observing only the requests corresponding to the files in the cache. We model this as a multi-armed bandit (MAB) \cite{MAB:Bubeck2012} problem,  and provide several caching algorithms. To the best of the authors' knowledge this is the first paper to address the optimal caching problem in the case of unknown popularity profiles using the MAB theory. The main contributions of the paper are:\myemph{\begin{itemize}
 \item We address the optimal content placement problem in an sBS when the popularity profiles of the available contents are not known in advance, while the instantaneous demands for files in the cache are observed periodically. 
 \item We show that this content placement problem is a MAB-problem, and propose three algorithms that efficiently learn the popularity profile and cache the best files.  
 \item We provide numerical results illustrating the impact of the system parameters on the performance of our algorithms.  
 \item We measure numerically the loss due to the lack of information about the popularity profile. 
 \item When the popularity profile is known, the problem reduces to the well-known knapsack problem, whose solution is known to be computationally hard. We propose a low-complexity  approximate algorithm to solve the knapsack problem, based on structural properties of its linear relaxation.
  \end{itemize}}

The rest of the paper is structured as follows:  the system model and the problem statement are presented in Section~\ref{sec:SM}. Section~\ref{sec:CMAB} studies the optimal caching problem when the popularity profile is not known. The optimal caching problem when popularity profile is known is addressed in Section~\ref{sec:off}.  Section~\ref{sec:res} presents extensive numerical results and, finally, Section~\ref{sec:con} concludes the paper.

\section{System Model}\label{sec:SM}

 We consider an sBS with a limited wireless backhaul link that provides high data rate service to its users. The sBS has a cache memory of capacity $M$ units, in which the CC can store files of its own choice. Time is divided into periods, and within each period there is a user request phase and a cache replacement phase. In the user request phase, users in the coverage area of the sBS request high data rate content. If a file is located in the sBS cache, a request is sent to the sBS, and the file is readily downloaded without incurring any cost to the macro cellular network; otherwise a request is sent to the macro BS. In the cache replacement phase, which is considered of negligible duration, the CC refreshes the cache content based on the content that has been downloaded so far, and the sBS broadcasts updated information about the  cache content to its users. Note that only requests served by the sBS are observed by the CC. 

 We denote by $\mathcal{F}$ the set of all files available,  by $F=|\mathcal{F}|$ the total number of files, and by $S_f$ the size of the $f$th file in $\mathcal{F}$. The set of all possible file sizes is $\mathcal{S}=\{s_1,\ldots,s_{|\mathcal{S}|}\}$, with $s_1>s_2>\ldots>s_{|\mathcal{S}|}$. We denote by $d^t_f$ the instantaneous demand for the $f$th file during user request phase in period $t$. The instantaneous demand, $d^t_f$, is an independent identically distributed (iid) random variable with mean $\theta_f$ and bounded support in $[0,U]$, where $U$ is the maximum number of users the sBS can serve at any given period. For simplicity, we assume identical file popularity over users, and independence among the demands of different users. Furthermore, we assume that the popularity profile of the files is characterized by a Zipf-like distribution with parameter $\gamma$, which is commonly used to model content popularity in networks \cite{caching:Breslau1999}. Hence, the popularity of file $f$, that is, the  expected number of requests in period $t$, is $\theta_f=\frac{U}{f^\gamma\sum_{i=1}^{F} \frac{1}{i^\gamma}}$, and the content popularity profile is $\pmb{\Theta}=(\theta_1,\ldots,\theta_F)$. Notice that parameter $\gamma$ models the skewness of the popularity profile. For $\gamma=0$ the popularity profile is uniform over files, and becomes more skewed as $\gamma$ grows. 
  
The  objective of the CC is to optimize the cache content at each time period in order to maximise the traffic served directly from the sBS, without knowing the popularity profile in advance, and by simply observing the requests corresponding to the files in the cache over time. 

A policy $\pi$ is an algorithm that chooses the cache content at each time period $t$, based on the whole history of the cached files and instantaneous demands. We denote by  $\mathcal{M}^t_\pi$ the set of files stored in the cache in period $t$, chosen according to $\pi$, and call it the \emph{cache content}. 
We consider a reward of $S_f$ units when file $f$ is fetched from the sBS. This reward can be considered as a QoE gain for the user, or a bandwidth alleviation on the macro cellular system.  We denote  the instantaneous reward associated with file $f$ by $r^t_f=d_f^t S_f $, and the expected instantaneous reward in period $t$ is 
\begin{equation} \label{eq:avrw}
  r_{\pmb{\Theta}}(\mathcal{M}^t_\pi)=\expected{\sum_{f\in \mathcal{M}^t_\pi }  d_f^t  S_f}{}=\sum_{f\in \mathcal{M}^t_\pi} \theta_fS_f,
\end{equation}
where the expectation is taken over the files' instantaneous demands. 
The objective is to find a policy $\pi$ that chooses  $\mathcal{M}^t_\pi$, $t=1,\ldots,N$,  which maximises the total expected accumulated reward over a time horizon $N$. This problem can be expressed as follows:
\begin{equation} 
\begin{aligned} \label{eq:opt_PG}
\max_{\pi}~ & ~ \sum_{t=1}^N  r_{\pmb{\Theta}}(\mathcal{M}^t_\pi) \\
&\sum_{f\in \mathcal{M}^t_\pi} S_f \leq M, ~~t=1,\ldots,N.
\end{aligned}
\end{equation}
If the  popularity profile, $\mathbf{\Theta}=(\theta_1, \ldots,\theta_F)$, is known, (\ref{eq:opt_PG}) can be divided into $N$ independent optimization problems, one for each period,  called the single-period optimization (SPO) problem. The SPO-problem is hard to solve, and is studied in Section \ref{sec:off}. Section~\ref{sec:CMAB} assumes the existence of an \mbox{($\mathbf{\alpha},\mathbf{\beta}$)-solver} for the SPO-problem, which is an algorithm that, with probability $\beta$, outputs a set of contents that provide at least $\alpha$ times the optimal reward.
Our main focus is on the more practically relevant case in which $\mathbf{\Theta}$ is not known in advance, and has to be estimated. Let $\hat{\mathbf{\Theta}}=(\hat{\theta}_1,\ldots,\hat{\theta}_F)$ denote the estimate of $\mathbf{\Theta}$. At each cache replacement phase, the popularity profile can be estimated based on the previous observations of the instantaneous rewards, and the SPO-problem can be solved using the  popularity profile estimate, $\hat{\mathbf{\Theta}}$. 
The instantaneous reward for files not cached in the sBS is not observed by the CC, and its estimate $\hat{\theta}_f$, for $f\notin\mathcal{M}^t_\pi$,  can not be updated in this period. This makes the problem more challenging, since the CC can obtain information on the popularity of a specific content only by placing this content in the cache. On the other hand the CC also wants to  exploit the limited storage capacity by caching the files that it believes to be the most popular. This is the well-known exploration vs. exploitation tradeoff. In Section \ref{sec:CMAB} we provide three algorithms to balance effectively these two factors.

\section{Learning the optimal cache content }\label{sec:CMAB}

\subsection{Multi-armed bandit problem}
The MAB problem \cite{MAB:Bubeck2012} models an agent with partial knowledge of the system. The agent takes actions repeatedly, based on its current knowledge, in order to maximise the accumulated reward overtime, while simultaneously acquiring new knowledge.    %It has been applied to a wide range of domains including online advertising \cite{MAB:W.Chen2013}, dynamic spectrum access \cite{Ahmad2009}, and energy harvesting \cite{P.Blasco2013}. 
The classic MAB problem considers a slot machine with $F$ arms, whose expected rewards are iid over time, with unknown means. At each time instant (e.g., the replacement phase in each period $t$ in our problem) one arm is pulled, and the slot machine yields a random reward.  The problem is to decide which arm to  pull at each time slot in order to maximise the  accumulated expected reward  over time.

The expected values of the arms are estimated based on  the instantaneous reward observations. The more times an arm is pulled the more reliable its estimate is, while the more times the arms with higher expected rewards are pulled the higher the expected accumulated reward is. Hence, there is a tradeoff between the exploration of new arms (i.e., pulling all arms a sufficient number of times to reliably estimate their mean rewards) and the exploitation of known arms (i.e., achieving higher rewards).

If the arms' expected rewards were known, the optimal policy would pull, at each time slot, the arm with the highest expected reward. The  \emph{regret} of a policy $\pi$ is the  difference between its expected accumulated reward and that of the policy that always pulls the best arm. Hence, the  regret is a measure of the loss due to not knowing the reward profile of the arms. The objective is to find a policy with a small regret.

In \cite{MAB:T.Lai1985}, Lai and Robbins show that no policy can achieve an asymptotic regret smaller than $O(\log(t))$, where $O(\cdot)$ includes a multiplicative and a constant term.   In \cite{MAB:P.Auer2002} a policy, called the upper confidence bounds (UCB), is presented, and proven to achieve a regret on the order of $O(\log(t))$, uniformly over~$t$.  

\subsection{MAB for optimal caching: Regret bound}

We will first illustrate that the problem presented in Section~\ref{sec:SM} is a generalisation of the classical MAB problem. Consider that each arm corresponds to a file in the system, that is, there is a total of $F$ arms, and that several arms can be pulled at the same time. At each period the CC decides which set of files are stored in the cache, which is equivalent to pulling a set of arms $\mathcal{M}^t_\pi$, and observes their instantaneous rewards. The instantaneous rewards are random variables with expected values $\theta_f S_f, \forall f \in \mathcal{M}^t_\pi$. With a slight abuse of notation, we denote by $\Theta$ the policy that is aware of the popularity profile, and that always caches the optimal content.  Let $R^\pi(t)$ denote the regret of policy $\pi$, i.e., $R^\pi(t)=r^{ac}_{\pmb{\Theta}}(t) - r^{ac}_\pi(t)$, where $r^{ac}_\pi(t)$ is the expected accumulated reward of policy $\pi$ until period $t$.  
A MAB problem in which several arms can be pulled simultaneously is known as a combinatorial MAB (CMAB) \cite{MAB:W.Chen2013} if the individual reward for each arm is observed, and as a linear MAB \cite{MAB:Abbasi-yadkori2011} if only an aggregated reward is observed. In this paper we consider the CMAB problem, since we assume that the individual instantaneous demands of all files in the cache are observed. An algorithm that has good theoretical results in terms of regret is the combinatorial UCB (CUCB) algorithm \cite{MAB:W.Chen2013}. Let $T_f$ denote the number of times file $f$ has been cached, $\hat{\theta}_f$ denote the instantaneous reward sample mean, and  $\bar{\theta}_f$ the perturbed version of $\hat{\theta}_f$. The specific embodiment of CUCB is given in Algorithm~\ref{alg:CUCB}. 

 \begin{algorithm}[h!]
\small\caption{\mbox{CUCB}} \label{alg:CUCB}
\begin{algorithmic}
   \STATE \textbf{$\mathbf{1.}$ Initialize:}
   \STATE  Cache all files at least once, observe the rewards, $r_f^t$, and update $\widehat{\theta}_f$ and $T_f$, $\forall f \in \mathcal{F}$.
   \STATE  Set  $t\leftarrow F+1$.
   \STATE \textbf{$\mathbf{2.}$  Observe (user request phase in period $t$):}
   \STATE  Observe $r_f^t$,  $\forall f \in \mathcal{M}^{t}_\pi$.
   \STATE  Update $\widehat{\theta}_f\leftarrow\frac{\widehat{\theta}_f\cdot T_f+r_f^t}{T_f+1}$, and $T_f\leftarrow T_f+1$, $\forall f \in \mathcal{M}^{t}_\pi$.
   \STATE Compute $\overline{\theta}_f=\widehat{\theta}_f+US_f\cdot\sqrt{\frac{3 \log (t) }{2 T_f}}$, $\forall f \in \mathcal{F}$.
   \STATE \textbf{$\mathbf{3.}$  Optimise (cache replacement phase in period $t$) :}
   \STATE Obtain $\mathcal{M}^{t+1}_\pi$ by using the \mbox{($\mathbf{\alpha},\mathbf{\beta}$)-solver}, with $\overline{\mathbf{\Theta}}=(\overline{\theta}_1, \ldots, \overline{\theta}_F)$. \vspace{-0.3cm}
   \STATE Cache files in $\mathcal{M}^{t+1}_\pi$.
   \STATE Set  $t\leftarrow t+1$.
   \STATE Go to step 2.
  \end{algorithmic}
\end{algorithm}

Notice that the algorithm does not use the estimates $\hat{\theta}_f$ to solve the SPO-problem, instead it uses the perturbed versions $\overline{\theta}_f$. The perturbation consists of an additive positive term, and its square grows logarithmically with $t$, and decreases linearly with $T_f$. The perturbation promotes files that are not placed often, by artificially increasing their expected reward estimates.    
Theoretical results in \cite{MAB:W.Chen2013} show that the regret of the CUCB algorithm is bounded by $O(\log(t))$. Notice that the regret bound does not converge. In theory the loss due to the lack of knowledge of the files' rewards grows to infinity. The power of this bound relies on the fact that, for large $t$ the loss grows only logarithmically with $t$.     The initial phase of the CUCB algorithm can be avoided by using some prior popularity estimates, for example, obtained from the content provider. 

\subsection{MAB for optimal caching: Application}

Despite the logarithmic growth of regret, CUCB can take many iterations to learn the optimal cache content. \myemph{In \cite{V.Kuleshov2000} it is shown for the classic MAB problem that, much simpler algorithms than UCB can achieve higher performance in practice.} One such simple algorithm is the \emph{\mbox{$\epsilon$-greedy}} algorithm, which caches at each iteration the best set of files according to the demand estimate $\hat{\mathbf{\Theta}}$ with probability $1-\epsilon$, and a random set of files with probability $\epsilon$. 

We propose yet another algorithm, based on CUCB, which we call the modified CUCB (MCUCB) algorithm. This algorithm exploits the Zipf-like distribution of the popularity profile, and the particular structure of the problem at hand. In MCUCB, the perturbation in step 2 of CUCB is modified as follows
\begin{equation}
     \overline{\theta}_f=\widehat{\theta}_f+\frac{U\cdot S_f}{F^\gamma} \sqrt{\frac{3 \log (Ut) }{2U T_f}},        
\end{equation}
where the factor $\frac{1}{F^\gamma}$ promotes exploitation when the Zipf distribution is skewed, that is, when $\gamma$ is large and there are few popular files. Parameter $\gamma$ can be empirically approximated as in \cite{caching:Breslau1999}. Exploitation is also promoted when $U$ is large, this reflects the fact that, in each period, $U$ independent realisations of the reward distribution are observed.  Numerical comparison of these three algorithms is relegated to Section \ref{sec:res}.

\section{\mbox{($\mathbf{\alpha},\mathbf{\beta}$)-solver} for the SPO-problem} \label{sec:off}
In this section we study the optimisation problem in (\ref{eq:opt_PG}) under the assumption that the popularity profile, $\mathbf{\Theta}$, is known. Since instantaneous rewards are iid random variables with known mean $\theta_f S_f$, the optimal cache content is independent of the period $t$, and (\ref{eq:opt_PG}) can be simplified by studying  a single period. We denote by $\mathbf{x}=(x_1,\ldots,x_f)$ the indicator vector of the files stored in the cache, where $x_f=1$, if file $f$ is in the cache, and $0$ otherwise.   The SPO-problem is:
\begin{equation} \label{eq:opt_off}
\begin{aligned}
\max_{\mathbf{x} } &~ r_{\pmb{\Theta}}(\mathbf{x}) =\sum_{f=1}^{F} \theta_f S_f x_f  \\
\text{s.t. } &
\sum_{f=1}^{F} S_f x_f \leq M \\
  & x_f \in \{0,1\}, ~f=1,\ldots, F. 
\end{aligned}
\end{equation}
 We denote by $\mathbf{x}^\Theta=(x_1^\Theta,\ldots,x_F^\Theta)$ the optimal cache content of the SPO-problem. We notice that (\ref{eq:opt_off}) has linear objective and constraint functions, while the optimisation variable is binary. In particular, (\ref{eq:opt_off}) is a knapsack problem with values $v_f=\theta_f S_f$, and weights $w_f=S_f$.  Knapsack problems, which are known to be NP-hard,  can be solved optimally using branching algorithms, such as branch and bound (\BAB) \cite{opt:Atamturk2005}. \BAB worst case complexity is exponential, same as exhaustive search. Approximate solutions  are obtained with low-complexity algorithms. 

\subsection{Approximate solution}
 We consider a linear-relaxation integer approximation of $\mathbf{x}^\Theta$ by relaxing the binary constraints on $x_i$ to $0\leq x_i \leq 1$.   Then (\ref{eq:opt_off})  becomes a linear program (LP), and can be solved in polynomial time. We denote by $\mathbf{x}^{LP}=(x_1^{LP},\ldots,x_F^{LP})$ the solution to the LP-relaxation.

 \begin{lem}\label{prop3}
  If  we reorder $\mathbf{\Theta}$ such that $\theta_i>\theta_j$ if $i<j$, $\forall i,j \in \mathcal{F}$, then the optimal solution has the following structure $\mathbf{x}^{LP}=(1,1,\ldots,1,\beta,0,0,\ldots,0)$, where $\beta=\sfrac{\left (M-\sum_{j=1}^{n-1}S_j \right)}{S_n}$, and $n$ is the coordinate of $\beta$ in $\mathbf{x}^{LP}$.
 \end{lem}

 If $\exists i,j \in \mathcal{F}$ such that $\theta_i=\theta_j$, there are several feasible solutions to the LP-relaxation problem, but at least one of those solutions fulfils  Lemma~\ref{prop3}. Since $\mathbf{x}^{LP}$ has a single non-integer element, we use its integer approximation to approximate the solution to (\ref{eq:opt_off}).  Let   $\mathbf{x}^{G}= \lfloor\mathbf{x}^{LP} \rfloor$ be an approximation to the optimal solution $\mathbf{x}^\Theta$, $\mathbf{x}^{G}$ is feasible, and only differ from $\mathbf{x}^{LP}$ in one element (i.e., the one equal to $\beta$ in Lemma~\ref{prop3}).  Due to the special structure of  $\mathbf{x}^{G}$, induced by Lemma~\ref{prop3}, it can be obtained with a \emph{greedy} algorithm, that caches  files sequentially, starting from the files with higher popularity, $\theta_f$, until the capacity of the cache has been reached. Hence, we call $\mathbf{x}^{G}$ the greedy approximation to $\mathbf{x}^{\pmb{\Theta}}$.
 
 We denote by $\delta=\frac{r_{\pmb{\Theta}}(\mathbf{x}^{\pmb{\Theta}})}{r_{\pmb{\Theta}}(\mathbf{x}^{G})}$, the ratio between the value of the optimal solution and that of the greedy approximation. If $\theta_i\geq \theta_j, \forall j<i$, and (\ref{eq:opt_off}) fulfils the \emph{regularity condition}, that is, $\sfrac{v_1}{w_1} \geq \ldots \geq \sfrac{v_F}{w_F}$, it can be estimated that $\delta \leq 2$ \cite{Korbut2010}. Hence, the greedy algorithm is an ($\alpha,\beta$)-solver with $\alpha=0.5$ and $\beta=1$.  For the general knapsack problem, the regularity condition ensures that Lemma \ref{prop3} hold \cite{Dantzig1957}. If $\mathbf{\Theta}$ characterises the Zipf distribution, we can obtain a tighter bound for $\delta$: 
 \begin{equation} \small
\delta\leq1+\frac{s_1}{s_{|\mathcal{S}|}}\frac{\sfrac{1}{{\left \lceil \frac{M}{s_1}\right \rceil}^\gamma}}{\sum_{i=1}^{\left \lfloor \frac{M}{s_1}\right \rfloor}\sfrac{1}{i^\gamma}} .
\end{equation}
If the cache size is much larger than the maximum file size, i.e., $M>>s_1$, then $\delta \approx 1$, and, hence, $\alpha\approx 1$.
\begin{rem}
For the special case of $|\mathcal{S}|=1$, i.e., when all the files have the same size, the solution of the greedy  approximation is optimal. 
\end{rem}
 
  \subsection{Optimal solution: \BAB}  
  
   In exhaustive search the objective function has to be evaluated for each point of the  feasible set. The \BAB algorithm discards some subsets of the feasible set without evaluating the objective function over these subsets, effectively reducing the solution time.
  \BAB iteratively creates branches by splitting the feasible set. Upper bounds for each branch are found using the LP-relaxation of the original problem. These upper bounds are then compared to the best lower bound found so far, and a branch is discarded if its upper bound is lower than the best lower bound, and it is further split otherwise. The initial lower bound on the optimal value can be obtained using the  greedy approximation. When the optimal solution of the LP-relaxation is feasible for the original problem (i.e., the solution is binary), the best lower bound is updated. The search stops when all branches have been discarded, and the best lower bound is the optimal value.

\begin{rem}
The greedy and \BAB algorithms, presented in this section, can be used as an \mbox{($\mathbf{\alpha},\mathbf{\beta}$)-solver} for the MAB algorithms of Section~\ref{sec:CMAB} (i.e., CUCB, MCUCB, and \mbox{$\epsilon$-greedy} MAB algorithms).
\end{rem}

\section{Numerical Results}\label{sec:res}
In this section the performances of the MAB algorithms presented in Section~\ref{sec:CMAB}, namely CUCB, MCUCB and \mbox{\mbox{$\epsilon$-greedy}}, are studied in an sBS cache system that provides high data rate service to its users.
First, the time evolution of these MAB algorithms is studied, and, for the \mbox{($\mathbf{\alpha},\mathbf{\beta}$)-solver}, the greedy approximation is compared to the optimal solution obtained with the \BAB algorithm. A number of numerical results involving different system parameters, such as the popularity profile, $\mathbf{\Theta}$, the number of sBS users, $U$, the cache memory size, $M$, and the total number of files in the system, $F$, will be presented. 

In addition to the MAB algorithms, an \emph{informed upper bound} (IUB) algorithm which assumes that the popularity profile is known in advance is also studied. The IUB algorithm provides an upper bound on the performance of any MAB algorithm. A well known algorithm for caching is the least recently used (LRU) algorithm. Each time a file is requested, and is not in the cache, LRU discards the least recently used file and caches the file that is requested. In our problem, since demands are observed only for those files in the cache, LRU is not applicable, and we consider the Myopic algorithm, which at each replacement phase keeps all the files that have been requested at least once within last user request phase and replaces randomly the rest of the files.  For comparison, we also consider the \emph{random} algorithm, which randomly caches files in the sBS until the cache memory is full. Note that no learning happens in the random algorithm, while the Myopic algorithm learns only from one-step past. 

We assume, unless otherwise stated, that $\mathcal{S}=\{1,3,5,7,9\}$ units, $U=100$, $\gamma=0.56$ (same as in \cite{caching:Golrezaei2012a} and \cite{caching:Breslau1999}), $F=1000$, $M=256$ units, and that there are $200$ files of each size. Notice that the cache memory can only store approximately $5.12\%$ of the total content size at any given time. In the rest of the paper, if the size of the cache is given in percentage, it is referred to the percentage of the total content size that can be stored in the cache memory at any given time. Finally, $\epsilon=0.07$ is used for the \mbox{$\epsilon$-greedy} algorithm.

The algorithms presented in Section~\ref{sec:CMAB} rely on the existence of an \mbox{($\mathbf{\alpha},\mathbf{\beta}$)-solver} in order to solve the SPO-problem. The accuracy of the greedy approximation proposed in Section~\ref{sec:off} is validated numerically. A total of $2\cdot10^4$ SPO-problems for different values of $F$, $M$ and $\mathbf{\Theta}$ are solved. According to our numerical simulations, the \BAB algorithm finds an optimal solution to the SPO-problem $95.5\%$ of the times within a $50$ seconds timeout. The greedy approximation is $99.99\%$ close the the optimal solution of the \BAB algorithm. As for the solution time, greedy algorithm is approximately $10^3$  times faster than the \BAB algorithm. In the rest of the paper, unless otherwise stated, the greedy approximation is used as the \mbox{($\mathbf{\alpha},\mathbf{\beta}$)-solver}.

\begin{figure}
  \centering
  \includegraphics[width=0.5\textwidth]{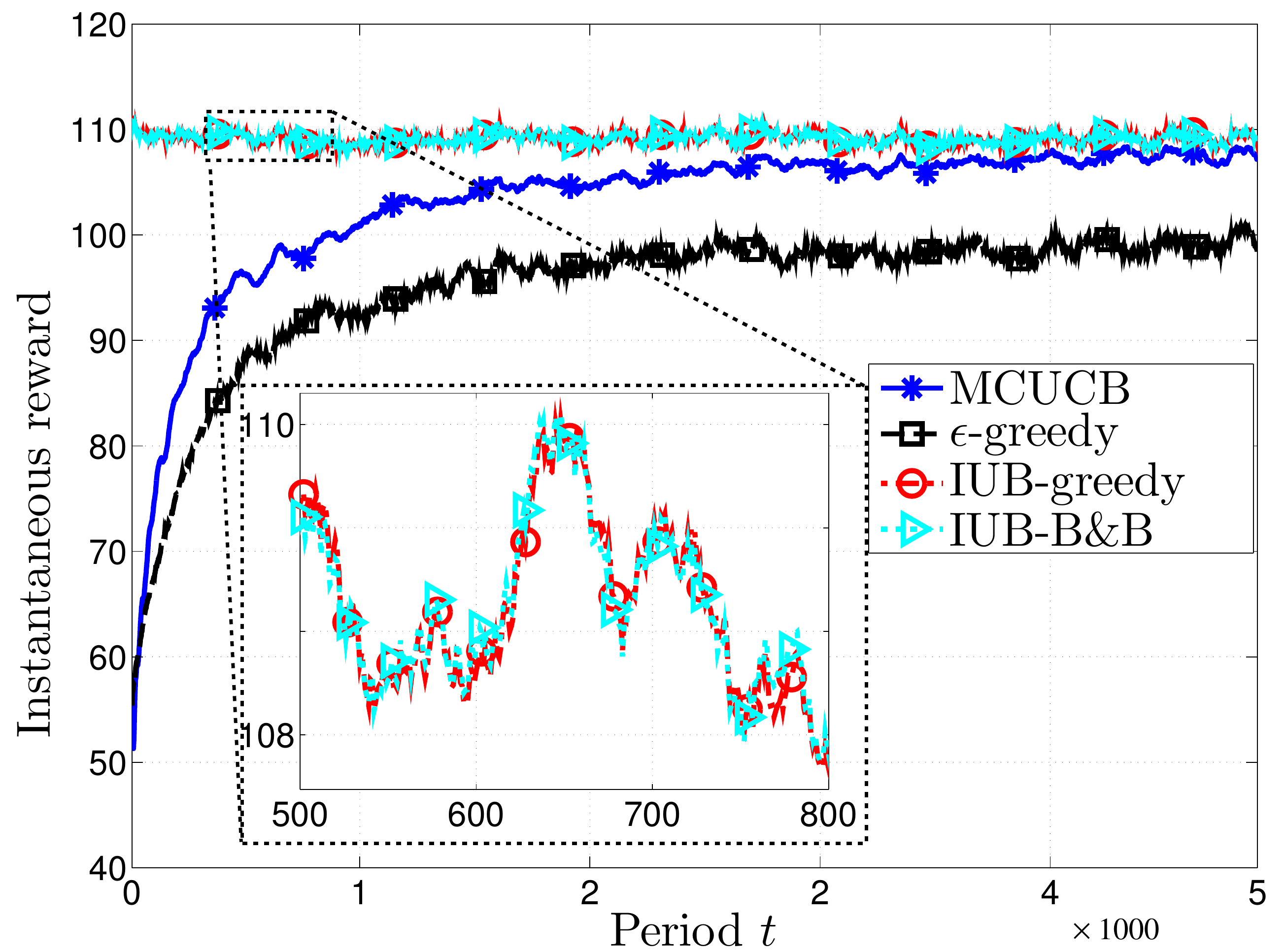}\\  \vspace{-0.3cm}
  \caption{The time evolution of the MCUCB and \mbox{$\epsilon$-greedy} algorithms, compared with the upper bound provided by the IUB algorithm (with both greedy and \BAB as the \mbox{($\mathbf{\alpha},\mathbf{\beta}$)-solver}).}\label{fig:conv2}\vspace{-0.5cm}
\end{figure}

\begin{figure}
  \centering
  \includegraphics[width=0.5\textwidth]{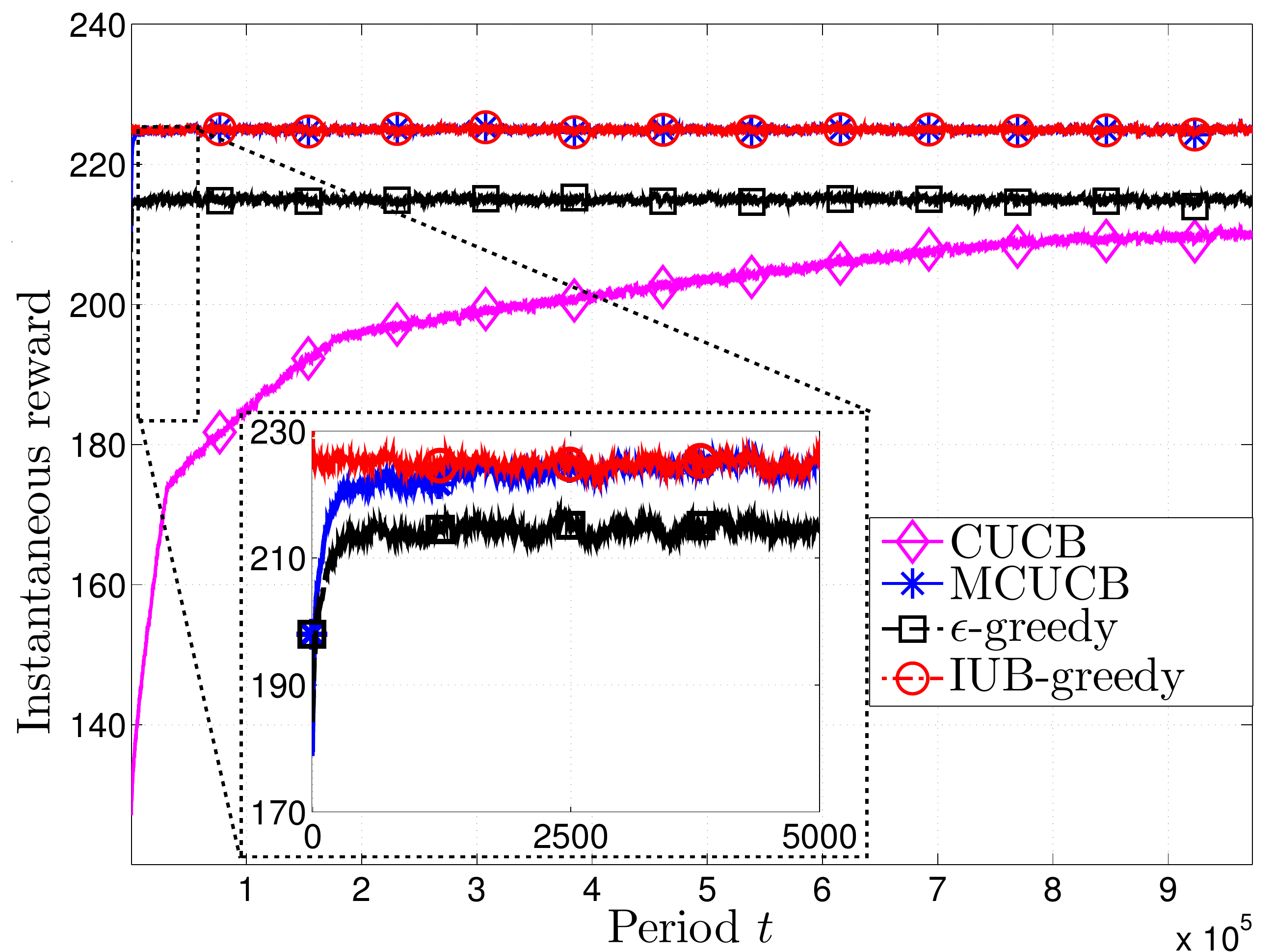}\\  \vspace{-0.3cm}
  \caption{Time evolution of the MAB algorithms, compared with the upper bounds provided by the IUB algorithms.}\label{fig:conv1} \vspace{-0.5cm}
\end{figure}

Time evolutions for the MCUCB and \mbox{$\epsilon$-greedy} algorithms are plotted in Figure~\ref{fig:conv2}.  The MCUCB algorithm achieves near optimal performance after 5000 iterations, whereas the \mbox{$\epsilon$-greedy} algorithm converges within a constant gap to the optimal performance of the IUB algorithm. This gap is due to the constant exploration factor $\epsilon$. For the IUB algorithm, we have considered the \BAB and the greedy approximation as the  \mbox{($\mathbf{\alpha},\mathbf{\beta}$)-solver}, and the difference between the two is insignificant. In our problem, the learning speed of the CUCB algorithm is very slow. In order to obtain results in a reasonable time frame, we reduce the dimension of the problem, i.e., $M=125$ and $F=100$. Figure~\ref{fig:conv1} shows the time evolution of the CUCB algorithm as well as that of the MCUCB and \mbox{$\epsilon$-greedy} algorithms. The MCUCB and \mbox{$\epsilon$-greedy} algorithms, despite the lack of theoretical convergence results, learn $100$ times faster than CUCB.

From this point onwards, we study the performance of the MCUCB and \mbox{$\epsilon$-greedy} algorithms after an initial learning stage of 2000 periods.  The performance is evaluated by the percentage of data that the users can fetch from the sBS cache memory, which is equivalent to the bandwidth alleviation in the macro cellular network. In the following, we present the performance of the algorithms averaged over $2\cdot10^4$ random realisations of the user requests.  

\begin{figure}
  \centering
  \includegraphics[width=0.5\textwidth]{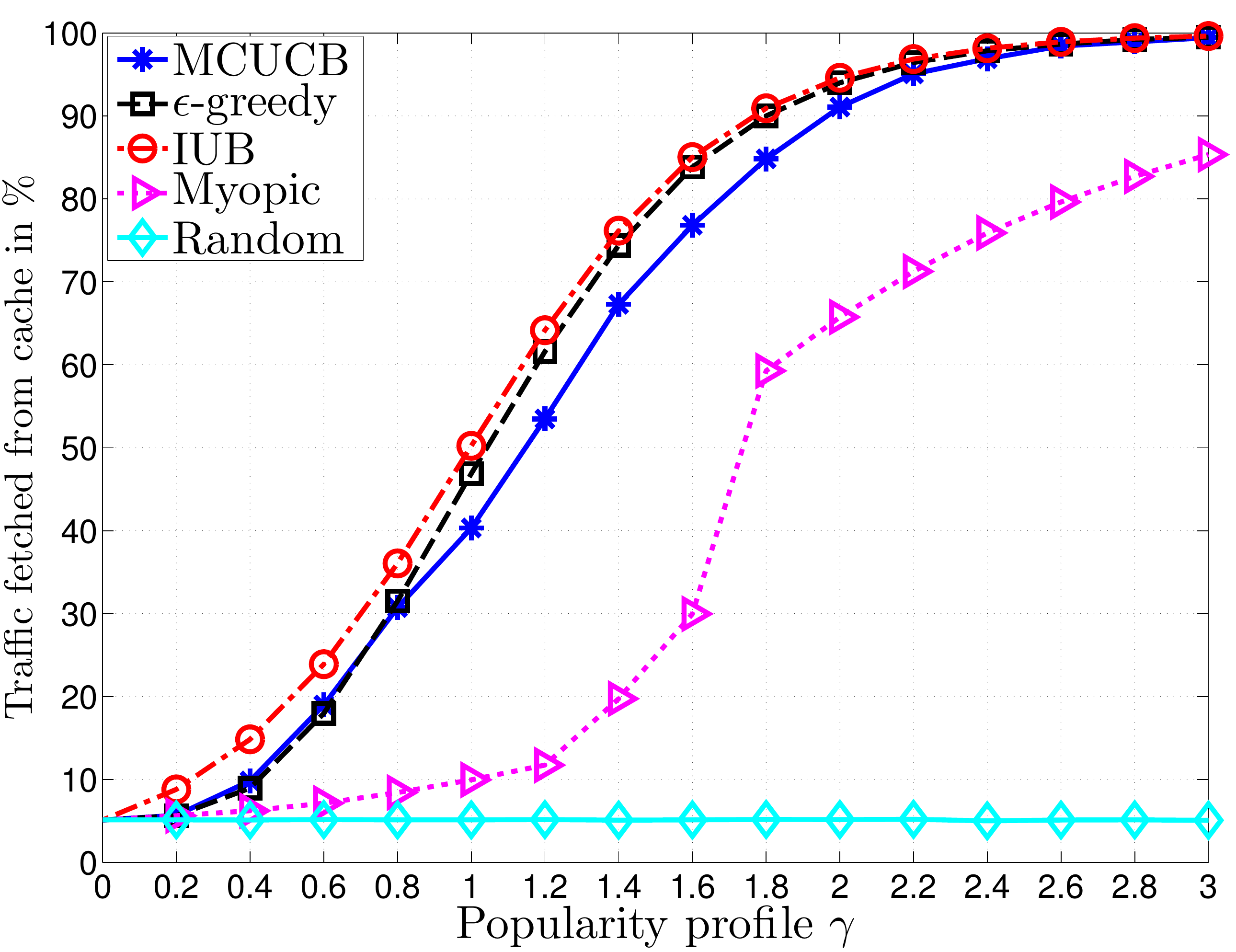}\\  \vspace{-0.3cm}
  \caption{Performance comparison of the MAB and IUB algorithms for different Zipf distribution parameters $\gamma$.}\label{fig:gamma}\vspace{-0.5cm}
\end{figure}

Figure~\ref{fig:gamma} shows the effect of the popularity profile on the performance. Clearly, when the popularity profile is uniform, that is when $\gamma$ is small, all algorithms have a performance close to $5\%$, that is the relative size of the cache memory. This is due to the fact that, if the demand is uniform the composition of the cache content is irrelevant. The random algorithm caches each period a random $5\%$-subset of the content total size, and achieves a  $5\%$ performance independent of the popularity profile.  As the popularity profile becomes more skewed, IUB upper bound, and the performances of the MCUCB and \mbox{$\epsilon$-greedy} algorithms increase until almost a $100\%$ of the users' demand can be fetched from the sBS's cache. Notice that the Myopic algorithm follows a similar trend, albeit much more slowly, and, theoretically, reaches the maximum performance when $\gamma$ approaches infinity.    

\begin{figure}  
  \centering
  \includegraphics[width=0.5\textwidth]{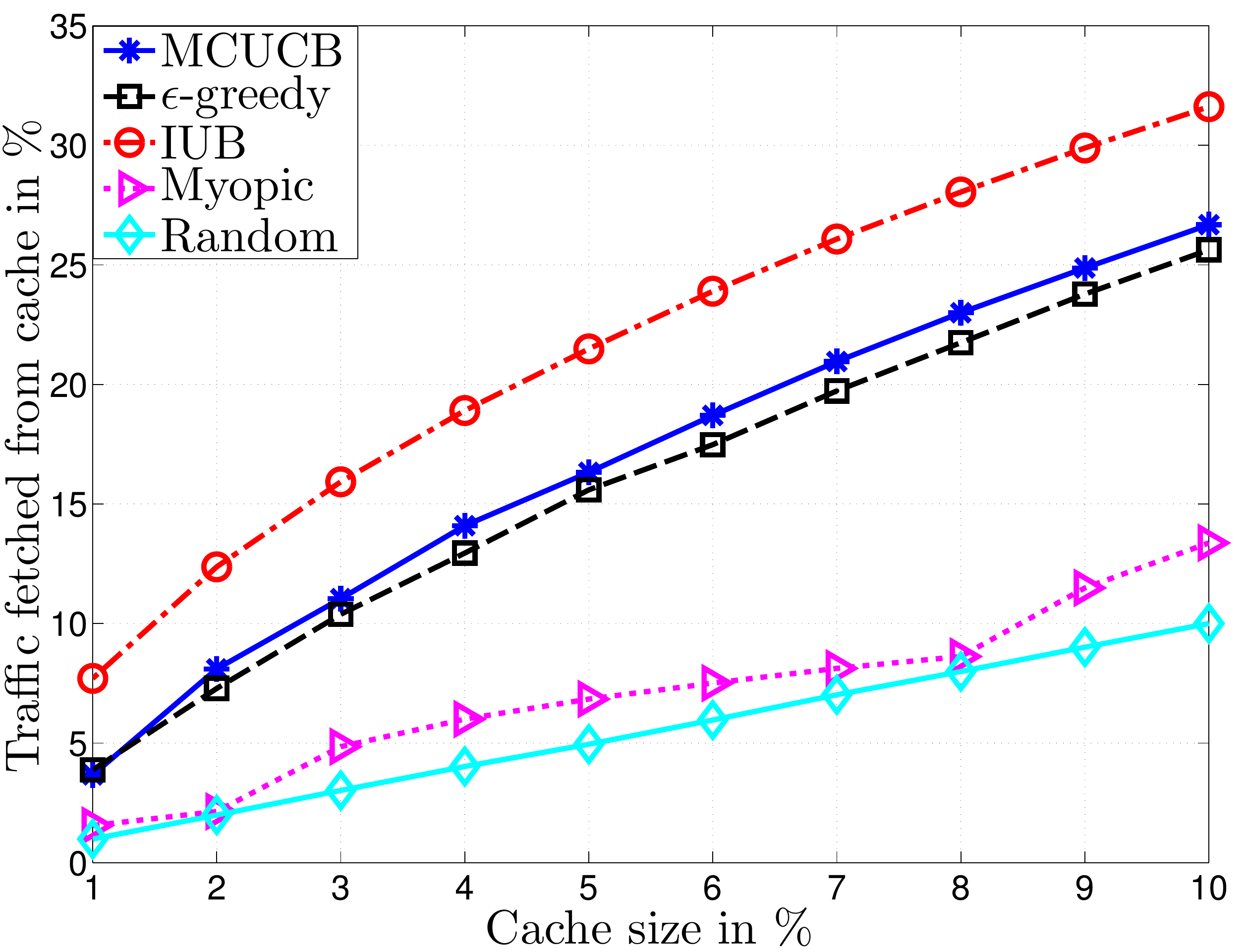}\\  \vspace{-0.3cm}
  \caption{Performance comparison of the MAB and IUB algorithms for different cache sizes, with $F=1000$.}\label{fig:cache}\vspace{-0.5cm}
\end{figure}

The performance with respect to the cache size, measured in percentage of the total content size, is studied in Figure~\ref{fig:cache}. As expected, performance of the random algorithm increases linearly with the cache size. Contrary to the results for $\gamma=[0.8, \ldots, 2.4]$ in Figure~\ref{fig:gamma},  in Figure~\ref{fig:cache} the MCUCB outperforms the \mbox{$\epsilon$-greedy} algorithm. This is because we set $\gamma=0.56$ in Figure~\ref{fig:cache}; and hence, the popularity profile is less skewed. The \mbox{$\epsilon$-greedy} algorithm estimates very accurately the popularity of the best files, whereas that of the files that are less popular is roughly estimated. For this reason, the more skewed the popularity is, the better the \mbox{$\epsilon$-greedy} algorithm performs. The MCUCB estimates the popularity of the files  with an accuracy more uniform compared to the \mbox{$\epsilon$-greedy} algorithm, which explains the gap between the two.

\begin{figure}
  \centering
  \includegraphics[width=0.5\textwidth]{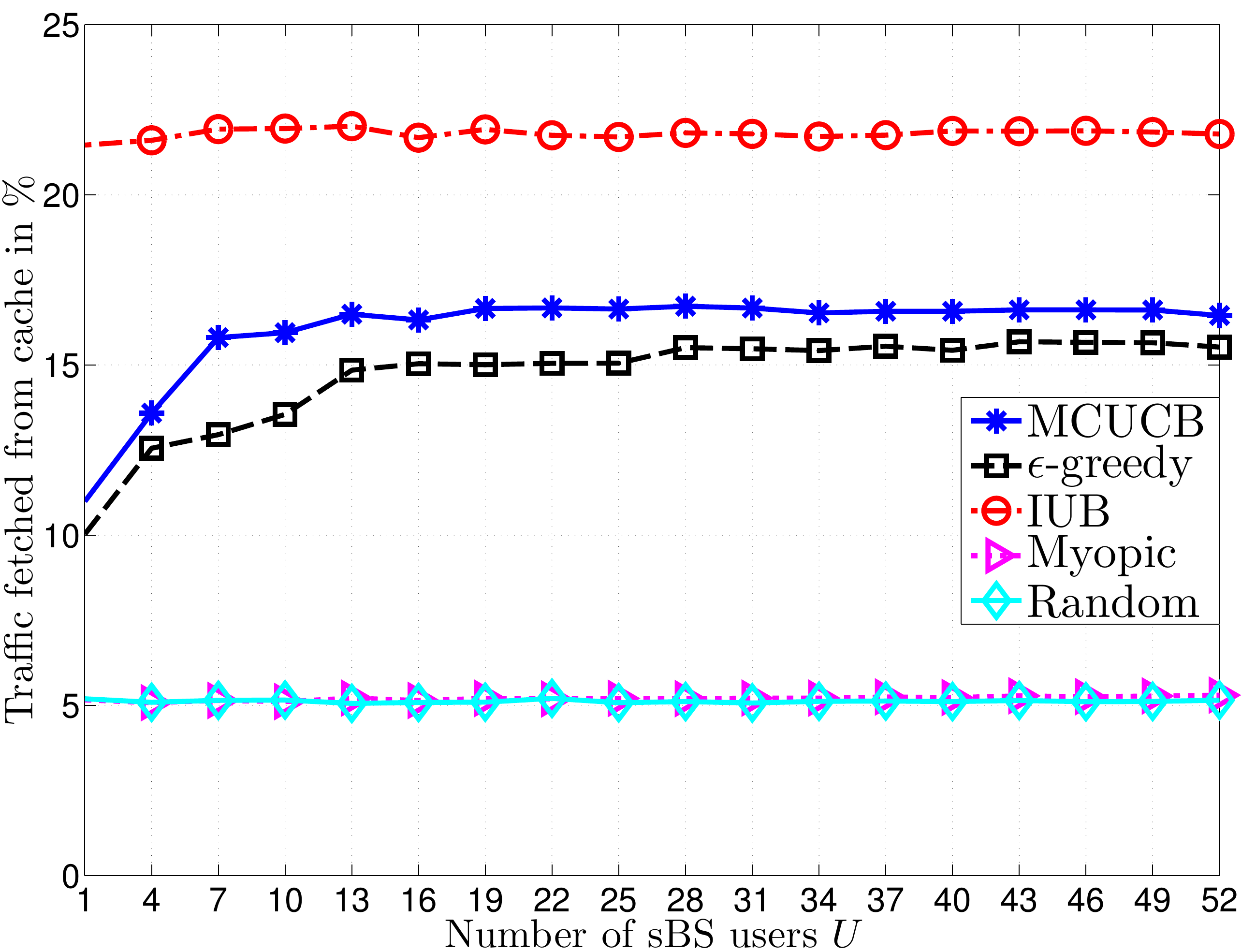}\\  \vspace{-0.3cm}
  \caption{Performance comparison of the MAB and IUB algorithms for different number of sBS users.}\label{fig:users} \vspace{-0.4cm}
\end{figure}

The performances of the MCUCB and \mbox{$\epsilon$-greedy} algorithms depend on the observations of the users' instantaneous demands. If the number of sBS users is low, the observations become less accurate, and both algorithms learn more slowly.  Figure~\ref{fig:users} depicts the performance of the proposed algorithms for different number of sBS users. When the number of sBS users is low, the performances of the MCUCB and \mbox{$\epsilon$-greedy} algorithms are reduced by $5\%$, whereas for more than 13 sBS users, the performances of both algorithms remain steady. 

\begin{figure}
  \centering
  \includegraphics[width=0.5\textwidth]{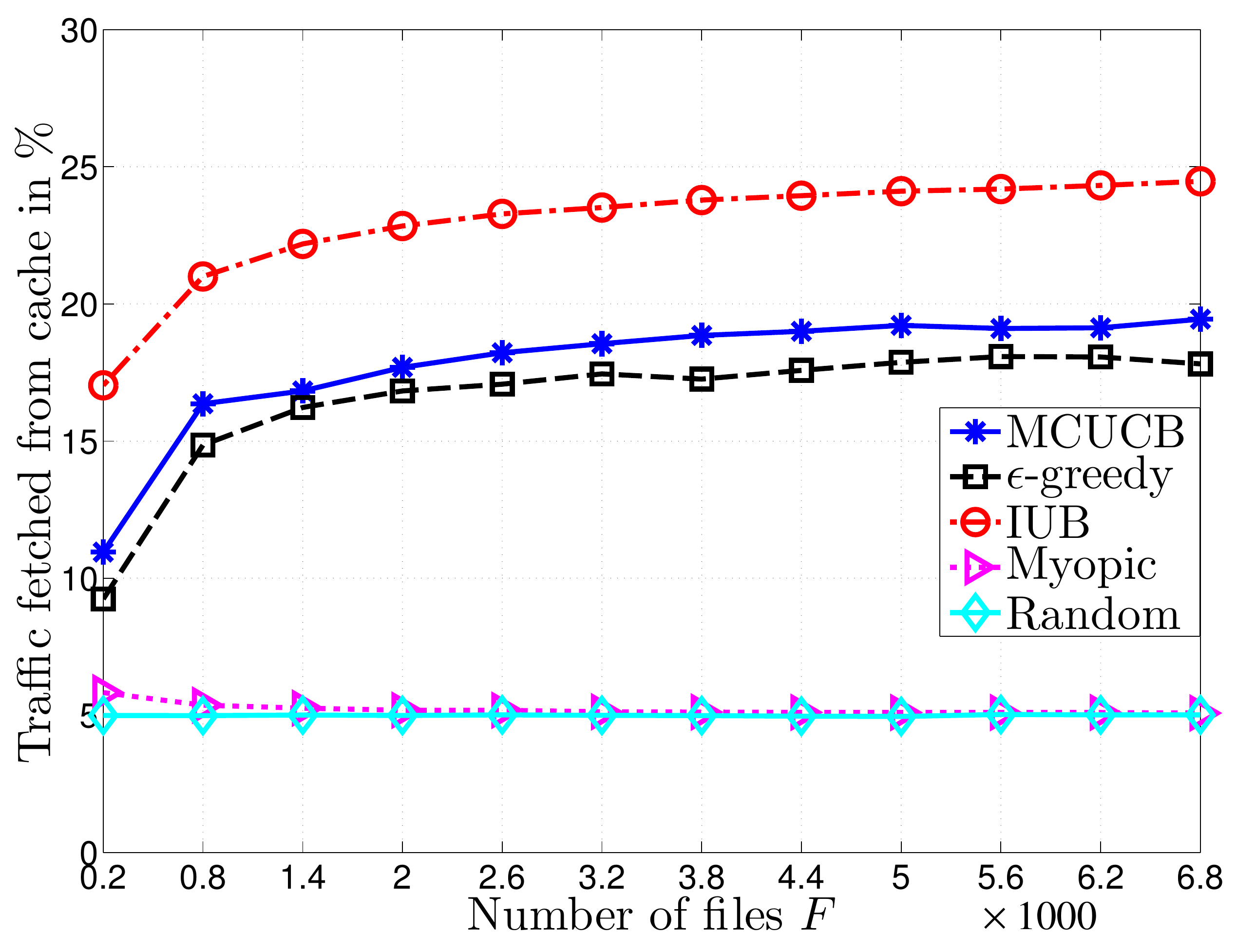}\\  \vspace{-0.3cm}
  \caption{Performance comparison of the MAB and IUB algorithms for different CP file set size $F$, with a fixed cache size of $5\%$.}\label{fig:file}\vspace{-0.5cm} 
\end{figure}

Finally, in Figure~\ref{fig:file} we study the effect of the number of files, $F$. The cache size is also changed such that, independent of $F$, the cache can always hold approximately $5\%$ of the content size. The popularity profile is more skewed for  large $F$, and  has a wider peak for  small $F$. Since the cache memory can store only $5\%$ of the files, when $F$ is small there are popular files that do not fit into the cache. The performance of the  IUB, MCUCB and \mbox{$\epsilon$-greedy} algorithms drop approximately by $5\%$  when $F$ is small, and grow steadily with $F$.  

 %\vspace{-0.2cm}
\section{Conclusions} \label{sec:con}
We have studied the optimal content caching problem in a backhaul-limited wireless sBS, when the file popularity profile is unknown and only the instantaneous demands of the files in the cache are observed. The sBS provides content selective offloading to its users. If a user wants to access high data rate content, and that content is in the sBS cache, it is fetched directly from the sBS, otherwise downloaded from the macro cellular network. Periodically, the CC optimises the cache content, based on the previous demands for the cached files,  so that the total traffic fetched from the sBS's cache memory is maximised. We have modeled the problem as a combinatorial MAB problem, and proposed several learning algorithms to optimise the cache content.  Theoretical results from the MAB literature upper bound the loss due to the lack of information about the popularity profile to grow logarithmically in time. 
We have observed numerically that, while the CUCB algorithm guarantees this theoretical bound, its performance in a practical system is quite poor. We have proposed an improved version, called the MCUCB, adapted to the demand distribution of the files, and showed that it performs significantly better than CUCB. Similarly, an \mbox{$\epsilon$-greedy} algorithm is also studied and showed to perform reasonably well for a wide range of system parameters. For the case studied in this paper in which only the instantaneous demands of the files stored in the cache are observed,  the MCUCB and \mbox{$\epsilon$-greedy}  algorithms outperformed a version of the well known LRU algorithm.

\vspace{-0.2cm}

\bibliographystyle{IEEEtran}
\bibliography{IEEEabrv,Totabiblio}

\flushend
\end{document}